# Keep an Eye on Venezuelan Elections!

Raúl Jiménez[1] and Manuel Hidalgo[2]

[1]Department of Statistics, Universidad Carlos III de Madrid

[2]Department of Political Science and Sociology, Universidad Carlos III de Madrid

**Abstract.** Starting with the 2004 recall referendum, an important opposition sector to President Chávez has questioned the integrity of the Venezuelan electoral system, and casts doubt on the legitimacy and impartiality of the upcoming 2012 presidential elections on October 7. After carrying out a forensic analysis on Venezuelan elections and referendums celebrated since 1998 until 2012, we reach two controversial conclusions: on one hand, we cannot rule out the hypothesis of fraud in elections run by the current regime. On the other hand, if fraud has been committed, this has hardly been decisive in the results of past elections. In other words, the winner would probably have been the same in clean elections. Only in a scenario of tight results, as 2012 elections could be, fraud would constitute a decisive factor.

The Law on Suffrage and Political Participation, approved in Venezuela in 1997 and reformed in 1998, establishes, as principle, the automatization of the vote count. During the elections and referendums in 1998, 1999, and 2000, the vote count was carried out both manually and in automatized form. However, since 2004, the totals exclusively come from a computer center where the results from the voting machines distributed in the country are centralized. Another characteristic that differentiates the electoral events before and after 2004 is the composition of the National Electoral Council (CNE), governing body of the Venezuelan elections. Its directorate, having been restructured due to the promulgation of the Law of the Electoral Power (2002), has taken controversial decisions that have only favored the government and never the opposition *(1)*. Moreover, if the current makeup of 4 to 1 *(2)* of the directorate is taken into account, it is quite natural that doubts arise about the integrity of the processes managed by this entity. Sectors from the opposition have reported fraud in various elections, regardless of the winner. The objective of this study is to evaluate these accusations in a unified manner through an innovative electoral forensic analysis. Unlike other methodologies, we not only detect outliers whose best explanation is an election-rigging *(3)*. Additionally, we test the hypothesis that atypical data may be the result of ad hoc measures, adopted to resolve issues that may have arisen on the voting day. The elections are complex processes, and, frequently, there are unforeseen events that may cause strong deviations from the expected distribution of voters in each polling station, which may entail false positives that must be discarded.

Despite the frequent use of the term, there is ambiguity regarding what is and what is not electoral fraud. What may constitute fraud in one country, or at a particular moment, may not be considered as such in another. Nonetheless, any action that is carried out with the intention of altering the development or material of an election, with the aim of affecting its results, may be considered a fraud *(4)*. Among the multiple irregularities that the opposition has denounced in the different elections, it is worth mentioning the hegemonic manipulation of the electoral registry, the coercion and intimidation by the public forces, and the use of technological platforms (voting machines and fingerprint scans) to cause distrust among the electorate. Such irregularities may add votes in favor of the governing party, and increase absenteeism to the detriment of the opposition. However, even though several of these denunciations have been proved, the amount of votes that have been compromised in the same is unknown. On some occasions, there have also been accusations of manipulating the vote count, the glaring violation of some vote centers, and the destruction of electoral material. How to prove that the irregularities being denounced have significantly affected the results? To this end, we propose the following sequential analysis:

1. Looking for traces of irregularities in the form of outliers, that is, results far removed from the expected ones. If we think like a CSI *(5)* in the electoral forensic analysis, this first step has to do with the search for marks left behind by the criminal, if there was one. We would like to stress that the presence of outliers (marks) is not an evidence of fraud. As mentioned before, electoral events are complex processes in which it is common to make mistakes and to take decisions along the way that may generate different types of atypical values.

2. Finding a strong correlation between the observed outliers and bias in the vote count that may result from irregularities significantly affecting the results. Resorting to the CSI metaphor again, this would be like determining that the traces found may be related to the crime.

3. Rejecting the hypothesis that the previous correlation has to do with factors different from the reported irregularities. This implies accepting that the irregularities do affect the results in order to favor one of the options. If this is like on TV, this would be where, in view of the evidence, the culprit confesses.

4. Estimating what would have been the results had no irregularities been committed. This is as close as the methodology can get to a crime scene reconstruction.

   We apply this sequence of steps to the following elections:

- Presidential elections, 1998, 2000, and 2006
- Referendums, 2004, 2007, and 2009
- Parliamentary elections 2010 (list vote)
- Presidential primary elections of the opposition, 2012

We did not consider the referendums of 1999 or the parliamentary elections of 2000 due to insufficient data at a level of breakdown that we required for our analysis. Due to the same reason, we did not consider the parliamentary elections of 2005, either, which were boycotted by the opposition by calling for abstention, foreseeing a fraud. Beside these exceptions, where the governing party won by a landslide, we have taken into account all the national elections since the implementation of the automatized vote in 1998.

**1. Detecting outliers that may have been caused by irregularities**. Let us call *innocent irregularity* any type of alteration in the expected distribution of voters that does not introduce a bias in the total counting of the votes. Even though it may seem an act of cheating at first for some, it is possible for irregularities reported by the opposition to be innocent, having affected both the opposition and governing party's forces without altering the electorate's will. With the intention of dismissing possible innocent irregularities a priori, we excluded a small number of centers reporting polling stations without votes. Such singularity may be attributed to failures in the system, and it is feasible that the voters of these stations have been directed to others. For the same reason, for the 2004 elections onward, we ruled out centers that included one or more stations where the ballot was cast manually. Even though these elections involve some non-automated centers, they are few and unusual, corresponding, in general, to remote locations with few voters. Lastly, for methodological reasons that are discussed at the end of this section, our analysis is limited to centers with two or more polling stations, which are the most. After this simple cleaning, there is a set of consistent data that accurately represents the official results of each election, which constitutes the base of our study.

Both for the Venezuelan case (*6,7*) and others (*8,9*), the reason for tracing irregularities in the form of atypical data associated to the numbers of abstentions, and void votes, has been justified. The Benford's test for the second significant digit (*10,11*) is one of the most extended tests of numeric anomaly of electoral data. Applying this test to the number of abstentions and void votes per station raises a loud alarm signal for the 2006 elections and 2009 referendum. On the contrary, Benford's law fits the 2000 and 2012 ones, as it should theoretically be if the numbers are not manipulated. For the remaining elections and referendums, the test provides evidence against this hypothesis. We summarize the obtained results on a table of *p*-values and a comparative plot between Benford's law and the observed frequencies (Table 1 and Fig. 1). Even though we recognize the usefulness of the Benford's test to alert about possible manipulations, we do not know how to filter possible false positives and false negatives that this test yields (*12*). An example of the former are the 1998 elections, which have been legitimized by all the international observers and political parties who participated in them (*13*); the low *p*-value that these elections yield are possibly due to innocent irregularities that the test does not discriminate.

| Año | 1998 | 2000 | 2004 | 2006 | 2007 | 2009 | 2010 | 2012 |
|---|---|---|---|---|---|---|---|---|
| *p*-valor | 0.0037 | 0.7915 | 0.0009 | 0 | $1.4 \times 10^{-5}$ | 0 | $5.3 \times 10^{-6}$ | 0.2518 |

**Table 1.** *p*-values of Benford's test for the second significant digit of the number of abstentions and void votes per polling station. The zero represents values below the numeric precision.

In this study, we detect outliers in the number of reported abstentions and void votes in a polling station through the *Z*-score

$$Z = \frac{O - p\tau}{\sqrt{p(1-p)\tau \frac{v - \tau}{v - 1}}}$$

where $O$ is the number of abstentions and void votes observed in the station, $p$ is the proportion of this number at the center to which the station belongs, $\tau$ is the number of voters registered at the station, and $v$ is the number of voters registered in the center. As it is discussed in (*6*), for stations in centers with two or more stations, the distribution of *Z*, conditioned to an absence of irregularities, is approximately distributed like a standard normal variable. The further *Z* is from zero, the further the number of abstentions and void votes per station are from the expected number. The normal probability plot of the *Z*-scores offers a qualitative measure of the joint behavior of the reported abstentions and void votes (Fig. 2). This simple analysis shows that all the processes present a clear pattern far from what is expected.

2. **Correlation between outliers and bias in the count.** Extreme values of *Z* may be due to:

   A. Random or innocent irregularities affecting an unbiased sample of stations, where the proportion of votes in favor of an electoral option over valid votes does not significantly differ from the population proportion.

   B. Innocent irregularities in a biased sample of stations.

   C. The alternative to *A* or *B*, that is to say, non-innocent irregularities affecting a set of stations, in favor of one of the electoral options.

Next, we discuss a test to prove the veracity of the following hypothesis:

*H1:* Extreme values of the Z-scores are due to *A*, only.

Let $\mathcal{M}_k$ represent the set of *k* stations with the *Z*-scores furthest from zero. We highlight that, regardless whether the cause is *A*, *B* o *C*, $\mathcal{M}_k$ is not a simple random sample; the stations in $\mathcal{M}_k$ have more or less votes than expected; and thus, we will find in them more variability than in those of a simple random sample. However, under *H1*, the proportion of votes in favor of one candidate over valid votes may not significantly differ between one sample and another. Let $r_k$ be the ratio between the number of votes in favor of Chávez, or his electoral proposals and supporters (Capriles, in the case of the 2012 primary elections), and the number of valid votes at the sample $\mathcal{M}_k$. Let *R* be the same ratio at all the polling stations considered in our study (*population ratio*). As we have discussed (*6*), the sample variance of the number of favorable votes per station is

$$s_k^2 = \frac{1}{k-1} \sum_{i \in \mathcal{M}_k} (W_i - r_k T_i)^2$$

where $W_i$ is the favorable votes on station *i* and $T_i$ is the total valid votes at the same station. Let us also consider the estimator of the variance of $r_k$

$$S_k^2 = \left(1 - \frac{k}{K}\right) \frac{1}{\mu_k^2} \frac{s_k^2}{k}$$

where *K* is the total population of stations and $\mu_k$ is the sample mean of valid votes per station (*14*). Standardizing, under *H1*, the distribution of

$$\zeta_k = \frac{r_k - R}{S_k}$$

can be approximated by a standard normal distribution for large values of $k < K - k$. Thus, extreme values of $\zeta_k$ for $100 < k < K/2$, are evidence against *H1*. We consider the series $\{\zeta_k, 100 < k < K/2\}$ of these statistics. If *K* is in the order of thousands, it is possible that some of these statistics fall out of the normal confidence interval of 99.99% $(-3.9, 3.9)$, even when *H1* is true. However, long excursions of the series away from this interval are highly improbable under *H1*, and constitute a convincing proof against this hypothesis.

With the aim of comparing the different processes, we calculated the $\zeta_k$ value for the case studies within the same range of *k* values, particularly for $100 < k < 1865$. The upper limit was chosen for being half the total number of stations of the elections with fewer stations (elections of 2000). The results that we present do not change if we analyze each process separately by making $k$ vary between 100 and $K/2$. The comparison (Fig. 3) allows visualizing the following groups:

- Referendums of 2004, 2007, and 2009, and elections of 2006, showing $\zeta_k$ values far from the normal confidence interval of 99.99% for a wide range of $k$ values.

- Elections of 1998, 2000, 2010, and 2012, with $\zeta_k$ values, always, or almost always, contained within the normal confidence interval of 99.99%.

We remark that in the elections and referendums of the first group (with the exception of 2007), we observe $p$-values, namely $P(|N(0,1)| > \zeta_k)$, under $10^{-6}$ (Fig. 4). Here, $N(0,1)$ represents a standard normal variable. On the contrary, in the elections of the second group (with the exception of 2010), many $\zeta_k$ values fall even within the normal confidence interval of 99% $(-2.58, 2.58)$. The significant bias in favor of Chávez and his proposals during the counting of votes at stations with extreme $Z$-scores values is a convincing evidence to reject *H1* in all the elections from the first group. However, we cannot reject this hypothesis in the elections from the second group.

**3. Are the outliers of 2004 and 2009 the result of innocent irregularities?** We have already discussed reasons (*6*) for the feasibility of the hypothesis

 *H2:*   Extreme $Z$-score values are due to *B*.

Regardless of what these reasons are, innocent irregularities on stations where the proportion of votes in favor of Chávez and his proposals significantly differs from the population proportion must be the result from the elections' own logistic and the electorate's idiosyncrasy. Therefore, if *H2* is true for the 2004 and 2009 referendums and elections, it should also be true for similar processes. In this paper, we have added evidence from 2010 to the ones already discussed (*6*) against this hypothesis. These elections were conducted by the same automatized system as the ones in the first group, showing that, similar to the 1998 and 2000 elections, the extreme $Z$-score values are not due to *B*, which lead us to dismiss *H2* for all the case studies. The only probable alternative explaining the observed irregularities between 2004 and 2009 is

 *H3:*   Extreme $Z$-score values are due to *C*.

At this point, a couple of reasonable questions to consider would be: If non-innocent irregularities were committed during the 2004, 2007, and 2009 referendums and the 2006 elections, why did Chávez's proposal lose in 2007? Why were not non-innocent irregularities committed in 2010? In order to answer them, we are going to explain the details of these two processes.

 The 2007 referendum on the constitutional reform was the first electoral defeat of Chavismo in nationwide elections. Although the opposition officially won by a narrow margin (roughly 1% of the votes), the definitive results are still unknown. The CNE announced the first bulletin with 86% of the polling stations processed, indicating that the tendency was "irreversible" (49% YES against 51% NO, with an abstention of 44%). Few days later, on a second bulletin, with 94% of the stations processed, the victory of NO was confirmed by leaving out the opinion of approximately 200,000 voters. The behavior of approximately 11% of the electoral census remains unknown, since the second bulletin presented total results, without indicating the results of all the stations. Such results have always raised questions and doubts. For example, while the initial percentage of abstention did not appear to have been calculated correctly, the percentage on the second bulletin is shocking (in average, the abstention rate is higher than 95% in the stations added). There have been different hypotheses about which option really won: from a larger victory of the opposition to a possible victory of Chavismo by a very narrow margin had all the votes been counted. We know that in the month of September 2007, surveys suggested the victory of YES but as weeks passed by the differences were shortened. In November, different opinion polls detected a trend change, where NO came to lead the surveys by 8-12 points, percentages that were reduced to a few points by the end of the month. This is why there was a talk of a dead heat. On the other hand, at the time of closing at the voting centers, some exit polls forecasted a tight result without a clear winner. However, quick counts based on a statistical sample, carried out by the NGO Súmate, estimated a difference between NO and YES greater than 8% (*15*). It pointed towards the direction of a possible manipulation of the vote count. The possibility that the governing entity had tried to make up the results with little room for maneuver may not be ruled out. In that case, the governing party had no other choice than to accept the victory of a NO under the risk of triggering a crisis of immeasurable consequences. The large mobilization of wide sectors of the opposition and its presence at the voting centers made the alteration of the final results difficult. Additionally, there was no consensus as to the reform within the governing party, which had a bearing on the demobilization of many supporters of Chávez. Lastly, according to the reported information, the Armed Forces made everyone respect the victory of the opposition.

During the parliamentary elections of 2010, the government resorted to another strategy. In light of a foreseeable decline of the electoral support reported by the surveys, it chose an electoral reform. Under the approved mixed-member majoritarian system, the percentage of deputies in the National Assembly, elected through plurality, increased from 60% to 70%. In addition, the reform legalized the practice of "morochas" ("twins", according to the Venezuelan idiom), with which the government's party had been clearly overrepresented since 2004 regional elections. With this practice, a party or coalition would create another ad hoc organization (*morocha*) representing a slate in the electoral district. Through the acquisition of a formal and independent status of this new organization, political parties avoided the subtraction of seats won through the plurality-majority system. In short, with the new electoral system known as parallel the election of the nominees in each slate is dissociated from the uninominal candidates. Thus, officialist candidates can take advantage of the high popularity of their leader. Without a doubt, the reform contravenes the principle of the personalization of suffrage in multi-member districts, as well as the Constitutional commitment to proportional representation (*16*). Finally, there were modifications in the electoral districts for a shameless gerrymandering; as it was made evident in the double strategy employed: isolating/concentrating zones that had voted against Chávez and his supporters in the past, and uniting areas with an electoral behavior favorable to the Government with other historically opposing ones. The governing party did very well out of the reform: even though the parties of the opposition as a whole obtained the majority of the votes, the ruling party won the absolute majority of seats in the National Assembly.

**4. Estimating results under the hypothesis of fraud.** The discussed methodology does not allow detecting stations that have been affected by irregularities if these do not generate atypical *Z*-score values. Hence, it is possible that the percentage of affected stations could be larger than what we are able to detect. Let this percentage (unknown) be denoted by $\beta \times 100\%$. The extreme case, with $\beta = 1$, would indicate that all the stations have been affected; $\beta = 0.5$ only 50% of them, etc. Let $\varepsilon$ be the difference (also unknown) between (a) the ratio between the number of votes in favor of Chávez or his proposals and the number of valid votes at the affected $\beta K$ stations, and (b) the same ratio if the stations had not been affected. Let $\rho$ be the *true population* ratio between the number of favorable votes and the number of valid votes, had not there been non-innocent irregularities; we only know the population ratio that the automatized vote count yields, which we have denoted as $R$. In case of fraud, if we know the values of $\beta$ and $\varepsilon$, we could estimate $\rho$ by $R - \beta\varepsilon$. Even though we cannot estimate $\beta$, we can estimate $\varepsilon$. If *H3* is true, the stations with extreme Z-scores values are, very possibly, a subsample of stations affected by irregularities; and hence, the difference between the ratio in this set of stations and the population ratio is a straightforward estimator of $\varepsilon$. In terms of the notation introduced above: Let $\kappa$ be the number of stations that have extreme Z-scores, less than $-3.9$ or greater than $+3.9$, for example (there are between 101 and 326 outliers of this kind, depending of the case study, between 34 and 60 times more than expected!). Then, the above estimator of $\varepsilon$ is $r_\kappa - R$. Consequently, a plug-in estimator of $\rho$ is $\rho_\beta = R - \beta(r_\kappa - R)$. We calculate this estimator for the elections and referendums as long as we have not disregarded the hypothesis of fraud. For every case, we consider possible different scenarios by varying $\beta$, and we compare them with the official results and other available data reported in the elections and referendums for which we cannot reject the *H3* alternative.

The recall referendum of 2004 has been widely debated *(6)*. Unlike the rest of the case studies that we have considered, only 150 stations were audited in it, and it is demonstrated that the sample is neither representative nor random *(17)*. Additionally, there is evidence that the counting of votes could have been altered for a high percentage of stations in the processing center *(18)*. By considering extreme scenarios in which 70% to 100% of the polling stations ($0.7 < \beta < 1$) were altered, our $\rho$ estimator oscillates around 0.49 and 0.51, pointing at a dead heat, which suggests that irregularities could have increased the chances of victory of the governing party, but could have hardly determined the winner. Some exit polls *(19)* suggest that the result was 61% in favor of the opposition, and not 59% in favor of the governing party, as reported by the CNE, but the confidence of these estimates are questionable *(6)*. Unfortunately, at least in the case of Venezuela, there are pre-electoral surveys and exit polls that suit all tastes.

Unlike the 2004 referendum, during the 2006 presidential elections, approximately 54% of the stations were audited and control over the automatized system was increased. With these guarantees, the opposition returned to the electoral battle after having called for the abstention in the 2005 parliamentary elections. It is agreed that the discrepancy discovered by the audits is not significant, thus ruling out the hypothesis of electronic fraud. But the scope of other irregularities affecting the results remains a mystery. Not even by considering an extreme scenario in which all the stations were affected by the irregularities, can we obtain estimates where the results are inverted. In fact, $\rho_1 = 0.56$, approximately 8% below the proportion reported by the CNE (62.84%). At most, we can support the statement of the defeated candidate (Manuel Rosales), who recognized his defeat by a margin that was narrower than what had been reported.

The referendum on the 2007 constitutional reform was discussed in detail in the previous section. Unlike the elections from 2006, the audits did reflect important differences between the casted and audited votes. According to the analysis carried out by Súmate, the difference in favor of the opposition was greater than 8%. Our estimates are compatible with this scenario for $\beta$ values close to 0.5, suggesting the manipulation of electoral results to dissimulate a clear defeat of the governing party, dissimilar to the near tie (50.7% versus 49.3%) decreed by the CNE.

The opposition admitted, albeit with deep reservations, the victory of the governing party during the 2009 referendum on the constitutional amend, where the nomination of President Chávez for the 2012 presidential reelection was approved. Similar to 2006, the audits matched the official results. However, they denounced, once again, the different irregularities mentioned before, unable to pinpoint the impact of the same. By assuming $β$ values between 0.25 and 0.45, we get estimates of $ρ$ that oscillate between 0.49 and 0.51, a dead heat. What happened in those elections is a mystery that our analysis cannot uncover. Quite possible, the official result of 55% in favor of the amend overestimates the electorate's will, but we cannot conclude that the irregularities determined the results.

Thus, even though we cannot dismiss the hypothesis of fraud during the elections managed by the current electoral referee, there has hardly been any that has been decisive in the results (*20*). Had it existed, it only could have possibly widened the difference of the results in favor of the governing party, decreasing the margin with which the opposition won in 2007. What will happen in the 2012 presidential elections if the results turn out to be as tight as they are predicted to be by some recognized surveys *(21)*? Our case studies suggest that in this scenario, the range of possible irregularities could overturn an election in favor of Chávez, if he were to end up being defeated.

**References and Notes:**

1. The Carter Center (2004) Executive summary of comprehensive report. Available: http://www.cartercenter.org/documents/1837.pdf

2. Among the five CNE directors, only one (Vicente Díaz) was appointed by the opposition. The rest have either been appointed by the governing party or held public offices under President Chávez.

3. Ehremberg, R. (2012). Election night numbers can signal fraud. *ScienceNews* 181:4:16.

4. López-Pintor, R. Assessing electoral fraud in new democracies: A basic conceptual framework. IFES Conference Paper, February 2010.

5. CSI (Crime Scene Investigation) is a famous television series on CBS.

6. Jiménez, R. (2011). Forensic analysis of the Venezuelan recall referendum. *Statist. Sci.* 26:4:564–583.

7. Levin, I., Cohn, G.A., Ordeshook, P.C., and Alvarez, R.M. Detecting voter fraud in an electronic voting context: An analysis of the unlimited reelection vote in Venezuela. EVT/WOTE'09 Proceedings of the 2009 conference on Electronic voting technology/workshop on trustworthy elections.

8. Myakgov, M., Ordeshook, P.C., and Shaikin D. (2009). *The Forensics of Election Fraud.* Cambridge University Press, New York.

9. Klimek, P., Yegorov, Y., Hanel, R., and Thurner, S. (2012). It's not the voting that's democracy, it's the counting: Statistical detection of systematic election irregularities. arXiv:1201.3087v1.

10. Mebane, W. (2008). Election forensics: The second-digit Benford's law test and recent American presidential elections. In Election Fraud: Detecting and Deterring Electoral

Manipulation. (R.M. Alvarez, T.E. Hall, and S.D. Hyde, Eds.) 162–181. Brooking Press, Washington, DC.

11. Pericchi, L. and Torres, D. (2011). Quick anomaly detection by the Newcomb–Benford Law, with applications to electoral processes data from the USA, Puerto Rico, and Venezuela. *Statist. Sci.* 26:4:513–527.

12. Deckert, J., Myagkov, M., and Ordeshook, P.C. (2010). The irrelevance of Benford's Law for detecting fraud in elections. *Political Analysis,* 19:3:245-268.

13. McCoy, J. (1999). Chávez and the end of "Partyarchy" in Venezuela. *Journal of Democracy* 10:3:64–77.

14. $S_k^2$ is a slight modification of the estimator used in (*6*), where we scale with the population mean rather than by $\mu_k$. Theory on ratio estimators is full discussed in Lohr, S. (2004). *Sampling: Design and Analysis, 2nd ed*. Brooks/Cole, Boston, MA.

15. Súmate (2008). *Informe Súmate. Referéndum sobre proyecto de reforma constitucional. Reporte de observación electoral.* Available online (in Spanish) at http://www.sumate.org.

16. Hidalgo, M. (2011). The 2010 legislative elections in Venezuela. *Electoral Studies*, 30:4:872–875.

17. Hausmann, R. and Rigobón, R. (2011). In search of the black swan: Analysis of the statistical evidence of fraud in Venezuela. *Statist. Sci.* 26:4:543–563.

18. Martín, I. (2011). 2004 Venezuelan presidential recall referendum (2004 PRR): A statistical analysis from the point of view of data transmission by electronic voting machines. *Statist. Sci.* 26:4:528–542.

19. Prado, R. and Sansó, B. (2011). The 2004 Venezuelan presidential recall referendum: Discrepancies between two exit polls and official results. *Statist. Sci.* 26:4:502–512.

20. Hidalgo, M. (2009). Hugo Chávez's "Petro-socialism". *Journal of Democracy*, 20:2:78–92.

21. Sagarzazu, I. (2012).Venezuelan Pollsters, their Records and the 2012 Race-August Update. Available online at http://venezuelablog.tumblr.com/

**Acknowledgements:** To Súmate, for having provided us the electoral data. R. Jiménez would like to thank the Department of Statistics and Operative Research of Universitat Politècnica de Catalunya, the Department of Mathematics of Lehigh University, and the Department of Mathematics and Statistics of the University of Portland, as well as their professors, for the opportunity to discuss the preliminary versions of this work, presented in different seminars during 2012. He is supported in part by Spanish MSI grant ECO2011-25706. M. Hidalgo is partially supported by Spanish MEC grant CSO2009-09233.

**Figures:**

**Fig. 1.** Benford's law for the second significant digit vs. observed frequencies of the number of abstentions and void votes per station.

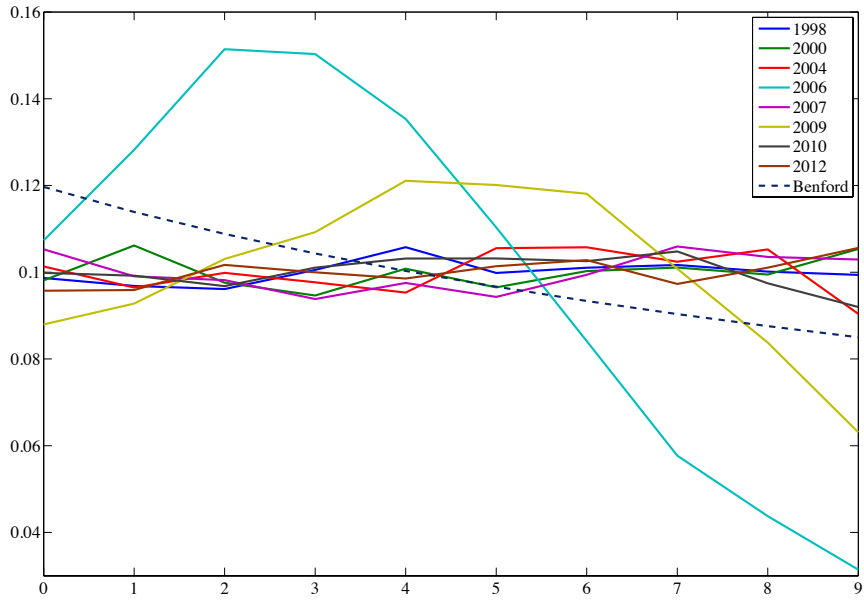

**Fig. 2.** Normal probability plots of $Z$-scores associated to the number of abstentions and void votes (+ = observed values and – - = expected values).

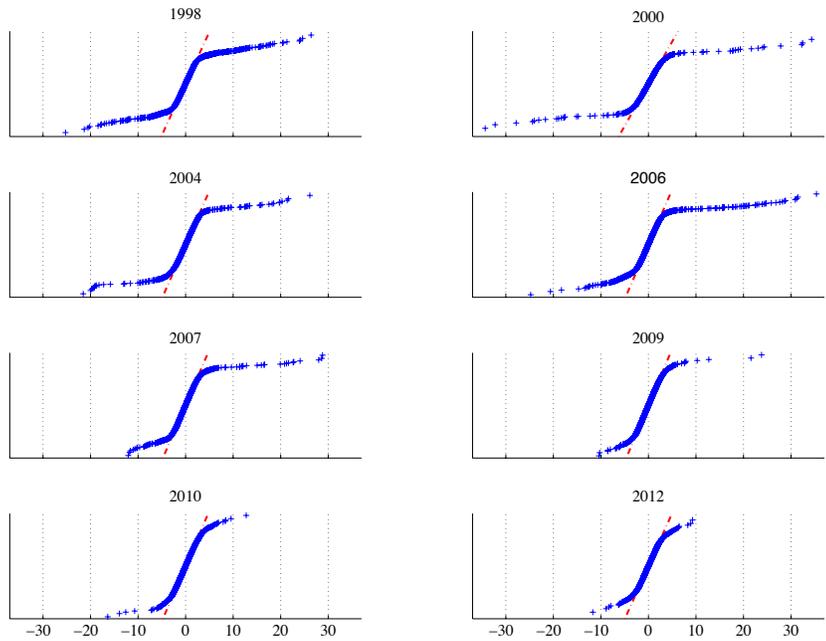

**Fig 3.** $\zeta_k$ versus $k$, for $100 < k < 1865$. The upper limit corresponds to half of the total stations being studied for the case with fewer stations (elections of 2000).

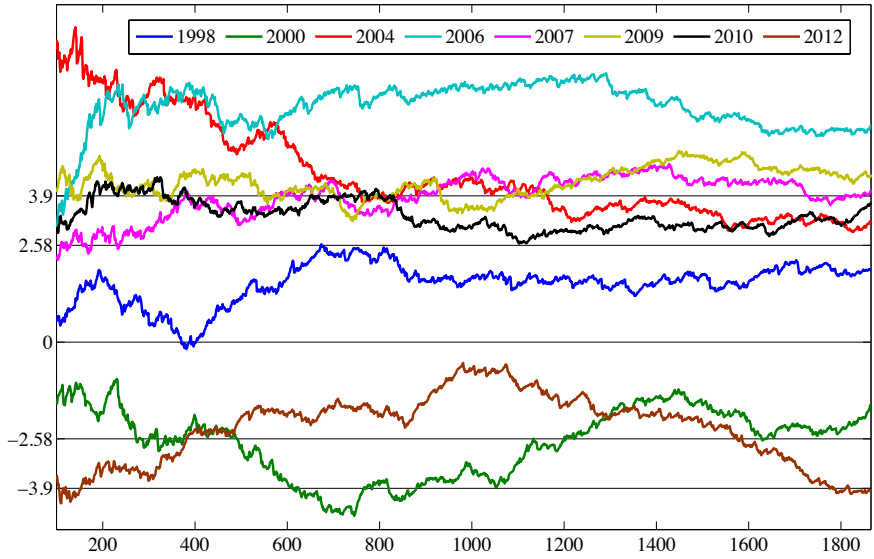

**Fig 4.** $p$-values $P(|N(0,1)| > \zeta_k)$ in logarithmic scale for $100 < k < 1865$.

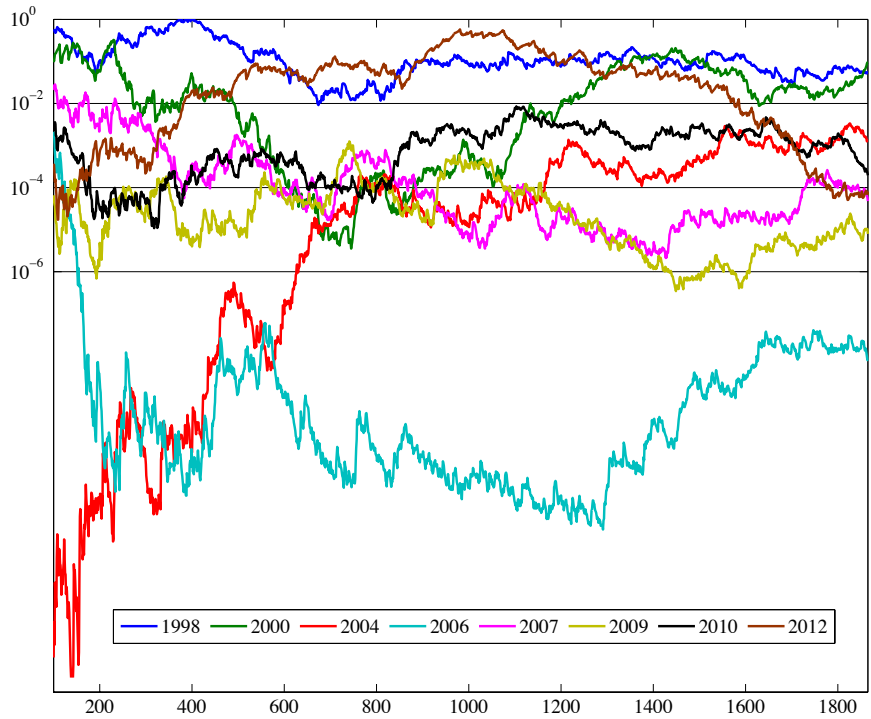

# ¡Vigilad las elecciones venezolanas!

Raúl Jiménez[1] y Manuel Hidalgo[2]

[1]Departamento de Estadística, Universidad Carlos III de Madrid

[2]Departamento de Ciencia Política y Sociología, Universidad Carlos III de Madrid

**Resumen.** A partir del referéndum revocatorio del 2004, un importante sector opositor al Presidente Chávez ha cuestionado la integridad del sistema electoral venezolano y tiene dudas acerca de la legitimidad e imparcialidad de las futuras elecciones presidenciales del 7 de Octubre del 2012. Practicando un análisis forense a elecciones y referéndums venezolanos desde 1998 hasta 2012 llegamos a dos controversiales conclusiones: por un lado, no podemos descartar la hipótesis de fraude en comicios administrados por el actual régimen. Por otro lado, de haberse cometido fraude en pasadas elecciones, difícilmente ha sido determinante en los resultados. Es decir, muy probablemente el ganador hubiese sido el mismo en elecciones limpias. Sólo en un escenario de resultados ajustados, como pudiera ser el 2012, el fraude podría ser determinante.

La Ley de Sufragio y Participación Política, aprobada en Venezuela en 1997 y reformada en 1998, establece como principio la automatización del escrutinio. En las elecciones y referéndums de 1998, 1999 y 2000 el conteo de votos se realizó tanto manualmente como en forma automatizada. Sin embargo, a partir del 2004 los totales son exclusividad de un centro informático, el cual centraliza los resultados de las máquinas de votación distribuidas en el país. Otra característica que diferencia los eventos electorales antes y después del 2004 es la composición del Consejo Nacional Electoral (CNE), organismo rector de las elecciones venezolanas. Su directorio, reestructurado a raíz de la promulgación de la Ley del Poder Electoral (2002), ha tomado decisiones controversiales que han favorecido sólo al gobierno, nunca a la oposición (*1*). Si a esto se le suma que la actual composición del directorio es 4 a 1 (*2*), es natural que surjan dudas acerca de la integridad de los procesos administrados por este ente rector. Sectores de la oposición han acusado fraude en varios comicios, independientemente de quien lo haya ganado. El objetivo de este trabajo es evaluar unificadamente estas acusaciones a través de un innovador análisis forense electoral. A diferencia de otras metodologías, nosotros no sólo detectamos datos atípicos cuya mejor explicación es un pucherazo (*3*). Adicionalmente, contrastamos la hipótesis de que estos valores puedan ser el resultado de medidas ad hoc tomadas para solventar problemas que se hayan podido presentar el día de la votación. Las elecciones son procesos complejos y frecuentemente ocurren imprevistos que pueden generar fuertes desviaciones de la distribución esperada de votantes por mesas, lo cual puede conllevar a falsos positivos que deben ser descartados.

A pesar del uso frecuente del término, existe ambigüedad acerca de qué es y no es fraude electoral. Lo que puede serlo en un país, o en un momento, puede no serlo en otro. No obstante, cualquier acción realizada con el propósito de alterar el desarrollo o el material de una elección con el fin de afectar los resultados puede ser considerado un fraude (*4*). Entre las múltiples irregularidades que la oposición ha denunciado en los diversos comicios destacamos la manipulación hegemónica del registro electoral, la coacción e intimidación por parte de los poderes públicos y el uso de plataformas tecnológicas (máquinas de votación y capta-huellas) para generar desconfianza en el electorado. Estas irregularidades pueden sumar votos a favor del oficialismo y aumentar la abstención, en detrimento de la oposición. Sin embargo, aunque varias de estas denuncias han sido demostradas, se desconoce la cantidad de votos involucrados en las mismas. En ocasiones, han llegado a denunciar la alteración del escrutinio de votos, la violación flagrante de algunos centros de votación y la destrucción de material electoral. ¿Cómo demostrar que las irregularidades denunciadas han afectado de manera significativa los resultados? Para ello proponemos el siguiente análisis secuencial:

5. Buscar rastros de las irregularidades en forma de datos atípicos. Esto es, resultados alejados de los esperados. Si pensamos en el análisis forense electoral como un CSI (*5*) en elecciones, este primer paso tiene que ver con la búsqueda de huellas dejadas por el criminal, si es que lo hubiere. Enfatizamos que la presencia de valores extremos (huellas) no es una prueba de fraude. Como dijimos anteriormente, los eventos electorales son procesos complejos en los que comúnmente se producen errores y se toman decisiones sobre la marcha que pueden generar diversos tipos de valores atípicos.

6. Encontrar una fuerte correlación entre los atípicos observados y sesgo en el conteo de votos que pueda ser producto de que las irregularidades afectan significativamente los resultados. Usando de nuevo CSI como metáfora, esto es como determinar que los rastros encontrados pueden estar relacionados con el crimen.

7. Rechazar que la correlación anterior se deba a factores distintos a las irregularidades denunciadas. Esto implica aceptar que las irregularidades sí afectan los resultados para favorecer a una de las opciones. Si las cosas fuesen como en la televisión aquí es donde, ante la evidencia, el culpable se confiesa.

8. Estimar cuáles hubiesen sido los resultados de no haberse cometido irregularidades. Esto es lo más parecido a la reconstrucción del crimen que la metodología puede ofrecer.

Aplicamos esta secuencia de pasos a las siguientes elecciones:

- Elecciones presidenciales 1998, 2000 y 2006
- Referéndums 2004, 2007 y 2009
- Elecciones parlamentarias 2010 (voto lista)
- Elecciones primarias de la oposición 2012

No consideramos los referéndums (consultivo y aprobatorio de la constitución) de 1999 ni las elecciones parlamentarias del 2000 por falta de datos al nivel de desagregación que requerimos para nuestro análisis. Por la misma razón, tampoco consideramos las elecciones parlamentarias del 2005, boicoteadas por la oposición, la cual llamó a abstención previendo un fraude. Quitando estas excepciones, donde las opciones oficialistas ganaron por mayoría abrumadora, hemos considerado todas las elecciones nacionales desde la implementación del voto automatizado en 1998.

**1. Detección de datos atípicos que han podido ser producto de irregularidades.** Vamos a llamar *irregularidad inocente* a cualquier tipo de alteración en la distribución prevista de votantes que no introduzca sesgo en el escrutinio total de votos. Aunque a priori le parezca a algunos una fullería, es posible que las irregularidades denunciadas por la oposición sean inocentes, que hayan afectado tanto a las fuerzas opositoras como a las oficialistas y que no hayan alterado la voluntad del electorado. Con la intención de descartar a priori posibles irregularidades inocentes, excluimos un pequeño número de centros que reportan mesas sin votos. Esta singularidad puede deberse a fallos del sistema y es factible que los votantes de estas mesas hayan sido redistribuidos a otras. Por la misma razón, para comicios del 2004 en adelante, descartamos centros que contengan una o más mesas en las que el voto se ejerció manualmente. Aunque en estos comicios existen algunos centros no automatizados, son pocos y peculiares, correspondiendo en general a localidades remotas con pocos electores. Por último, por razones metodológicas que se discuten al final de esta sección, nuestro análisis se restringe a centros con dos o más mesas electorales, que son la mayoría. Después de esta simple limpieza se obtiene un conjunto de datos consistente que representa fielmente los resultados oficiales de cada elección y que es la base de nuestro estudio.

Tanto para el caso venezolano (*6,7*) como para otros (*8,9*) se ha justificado por qué rastrear irregularidades en la forma de datos atípicos asociados al número de abstenciones y votos nulos. El test de Benford para el segundo dígito significativo (*10,11*) es una de las pruebas más difundidas de anomalía numérica de datos electorales. La aplicación del test al número de abstenciones y votos nulos por mesa dispara una estridente señal de alarma para la elección del 2006 y el referéndum del 2009. Por el contrario, no rechaza que el 2000 y el 2012 ajusten la ley de Benford, como en teoría debe ocurrir si los números no son manipulados. Para el resto de elecciones y referéndums, el test aporta evidencias en contra de esta hipótesis. Resumimos los resultados obtenidos con una tabla de *p*-valores y un gráfico comparativo entre la ley de Benford y las frecuencias observadas (Tabla 1 y Fig. 1). Si bien reconocemos la utilidad del test de Benford para alertar sobre posible manipulaciones, no sabemos como descartar falsos positivos y falsos negativos que este test podría arrojar (*12*). Un ejemplo de los primeros son las elecciones de 1998, que han sido legitimadas por todos los observadores internacionales y actores políticos que participaron en ellas (*13*); posiblemente el bajo *p*-valor que arrojan estas elecciones se deba a irregularidades inocentes que el test no discrimina.

| Año | 1998 | 2000 | 2004 | 2006 | 2007 | 2009 | 2010 | 2012 |
|---|---|---|---|---|---|---|---|---|
| *p*-valor | 0.0037 | 0.7915 | 0.0009 | 0 | $1.4\times10^{-5}$ | 0 | $5.3\times10^{-6}$ | 0.2518 |

**Tabla 1.** *p*-valores del test de Benford para el segundo dígito significativo del número de abstenciones y votos nulos por mesa. El cero representa valores por debajo de la precisión numérica.

En este trabajo medimos la atipicidad del número de abstenciones y votos nulos reportados en una mesa mediante el *Z*-score

$$Z = \frac{O - p\tau}{\sqrt{p(1-p)\tau\frac{v-\tau}{v-1}}}$$

Siendo $O$ el número observado en la mesa, $p$ la proporción de abstenciones y votos nulos en el centro al cual pertenece la mesa, $\tau$ el número de electores registrados en la mesa y $v$ el número de electores registrados en el centro. Como se discute en (*6*), para mesas en centros con dos o más mesas, la distribución de *Z* condicionada a que no hay irregularidades se distribuye aproximadamente como una normal estándar. En la medida que *Z* está más lejos de cero el número de abstenciones y nulos de la mesa está más alejado de lo esperado. El gráfico de probabilidad normal de los Z-scores ofrece una medida cualitativa de la atipicidad conjunta de las abstenciones y votos nulos reportados (Fig. 2). Este simple análisis muestra que todos los procesos presentan un claro patrón lejos de lo esperado.

**2. Correlación entre atípicos y sesgo en el escrutinio.** Valores extremos de $Z$ pueden deberse a:

D. Azar o irregularidades inocentes que afectan una muestra insesgada de mesas, donde la razón de votos a favor de un candidato entre votos válidos no difiera significativamente de la razón poblacional.

E. Irregularidades inocentes en una muestra sesgada de mesas.

F. La alternativa a $A$ o $B$, es decir a irregularidades no inocentes que afectan un conjunto de mesas a favor de una de las opciones electorales.

A continuación discutimos un test para probar la veracidad de la siguiente hipótesis:

*H1:*   Valores extremos de los $Z$-scores se deben sólo a $A$.

Denotemos por $\mathcal{M}_k$ al conjunto de las $k$ mesas con $Z$-scores más alejados de cero. Resaltamos que, sin importar si la causa es $A$, $B$ o $C$, $\mathcal{M}_k$ no es una muestra aleatoria simple; las mesas de $\mathcal{M}_k$ tienen más o menos votos de lo esperado y por consiguiente encontraremos en ellas más variabilidad que en los de una muestra aleatoria simple. Sin embargo, bajo *H1*, la razón de votos a favor de un candidato entre votos válidos no debe diferir significativamente entre una u otra muestra. Denotemos por $r_k$ la razón entre votos favorables a Chávez y sus propuestas (a Capriles en el caso de las primarias del 2012) y total de votos válidos en la muestra $\mathcal{M}_k$ y por $R$ la razón poblacional. Como ya hemos discutido (*6*), la varianza muestral del número de votos favorables por mesa es

$$s_k^2 = \frac{1}{k-1} \sum_{i \in \mathcal{M}_k} (W_i - r_k T_i)^2$$

siendo $W_i$ los votos favorables en la mesa $i$ y $T_i$ el total de votos válidos en la misma mesa. Consideremos también el estimador de la varianza de $r_k$

$$S_k^2 = \left(1 - \frac{k}{K}\right) \frac{1}{\mu_k^2} \frac{s_k^2}{k}$$

donde $K$ es el total poblacional de mesas y $\mu_k$ es el promedio muestral de votos válidos por mesa (*14*). Estandarizando, bajo *H1*, el estadístico

$$\zeta_k = \frac{r_k - R}{S_k}$$

se distribuye aproximadamente como una normal estándar para valores grandes de $k < K - k$. Así, resultados extremos del estadístico $\zeta_k$, digamos para $100 < k < K/2$, son evidencias contra *H1*. Nosotros consideramos la serie $\{\zeta_k, 100 < k < K/2\}$ de estos estadísticos. Si $K$ es del orden de miles, es posible que algunos de estos estadísticos caigan fuera del intervalo de confianza normal del 99.99% $(-3.9, 3.9)$, aún cuando *H1* sea cierta. Sin embargo, largas excursiones de la serie alejadas de este intervalo son muy improbables bajo *H1* y constituyen una prueba convincente contra esta hipótesis.

Con el objeto de comparar los distintos procesos, calculamos el valor de $\zeta_k$ para todo el estudio de casos dentro del mismo rango de valores de $k$, en concreto para $100 < k < 1865$. El límite superior fue escogido por ser la mitad del total de mesas de las elecciones con menos mesas (elecciones del 2000). Los resultados que presentamos no cambian si analizamos por separado cada proceso haciendo variar $k$ entre 100 y $K/2$. La comparación (Fig. 3) permiten visualizar el siguiente agrupamiento:

- Referéndums del 2004, 2007 y 2009 y elecciones del 2006, que muestran valores de $\zeta_k$ alejados del intervalo de confianza normal del 99.99% para un amplio rango de valores de $k$.

- Elecciones de 1998, 2000, 2010 y 2012, con valores de $\zeta_k$ siempre, o casi siempre, contenidos en el intervalo de confianza normal del 99.99%.

Resaltamos que en las elecciones y referéndums del primer grupo (con la excepción del 2007) se observan *p*-valores, a saber $P(|N(0,1)| > \zeta_k)$, por debajo de $10^{-6}$ (Fig. 4). Aquí $N(0,1)$ representa una variable normal estándar. Por el contrario, en las elecciones del segundo grupo (con la excepción del 2010) una buena parte de valores de $\zeta_k$ caen incluso dentro del intervalo de confianza normal del 99% $(-2.58, 2.58)$. El significativo sesgo a favor de Chávez en el conteo de votos en mesas con valores extremos de $Z$ son una evidencia contundente para rechazar *H1* en todos los comicios del primer grupo. Por el contrario, el mismo análisis no aporta evidencias para rechazar la hipótesis en los comicios del segundo grupo.

### 3. ¿Son las atipicidades observadas del 2004 al 2009 producto de irregularidades inocentes?

Ya hemos discutido (*6*) razones que explicarían la factibilidad de la hipótesis

*H2:*       Valores extremos de los *Z*-scores se deben a *B*.

Sin importar cuales sean estas razones, irregularidades inocentes en mesas donde la razón de votos a favor de Chávez difiera significativamente de la poblacional deben ser resultado de la propia logística de los comicios y de la idiosincrasia del electorado. Por ende, si *H2* es cierta para los referéndums y elecciones del 2004 al 2009 también debe serlo para procesos similares. A las evidencias ya discutidas (*6*) en contra de esta hipótesis hemos añadido en este trabajo las del 2010. Estas elecciones fueron conducidas con el mismo sistema automatizado que las del primer grupo y demuestran, al igual que las elecciones de 1998 y 2000, que los valores extremos de los *Z*-scores no se deben a *B*, lo que nos lleva a descartar *H2* para todos los estudio de casos. La única alternativa probable que explica las irregularidades observadas entre el 2004 y el 2009 es

*H3:*       Valores extremos de los *Z*-scores se deben a *C*.

La pregunta natural que surge ahora es: ¿si se cometieron irregularidades no inocentes en los referéndums del 2004, 2007 y 2009 y en las elecciones del 2006, por qué perdieron en el 2007 y por qué no las cometieron en el 2010? Para responderlas, vamos a explicar detalles de estos dos procesos.

En el Referéndum de la Reforma Constitucional del 2007 se produjo la primera derrota electoral del chavismo en unas elecciones nacionales. Aunque oficialmente ganó la oposición por un estrecho margen (poco más del 1% de los votos), todavía no se conocen los resultados definitivos. El CNE anunció los resultados con el 86% de mesas totalizadas indicando que la tendencia era "irreversible" (49% del SI frente al 51% del NO con una abstención del 44%). Días después, en un segundo boletín, con el 94% de mesas totalizadas, se confirmó el triunfo del NO, excluyendo la opinión de aproximadamente 200.000 votantes. Tales resultandos siempre han suscitado interrogantes y sospechas ya que, entre otras cuestiones, se desconoce el comportamiento de aproximadamente el 11% del censo electoral (en el segundo boletín se presentaron los resultados globales, sin detallar los resultados de todas las mesas). Además, mientras que el porcentaje de abstención inicial no pareciera haberse calculado de manera correcta, el presentado en el segundo boletín es chocante (en promedio, la abstención superó el 95% en las mesas añadidas). Distintas hipótesis se han barajado sobre cuál opción ganó en realidad: desde un triunfo más amplio de la oposición hasta una posible victoria del chavismo por un margen muy estrecho, de haberse contado todos los votos. Sabemos que en el mes de septiembre de 2007 las encuestas daban un triunfo del SI pero conforme pasaron las semanas las diferencias se acortaron. En el mes de noviembre diversos estudios demoscópicos detectaban un cambio de tendencia, pasando el NO a encabezar las encuestas en 8-12 puntos, porcentajes que se redujeron a unos pocos puntos a finales de dicho mes, de ahí que se llegara a hablar de un empate técnico. Por otro lado, al momento de cierre de los centros de votación, algún *exit poll* vaticinaba un resultado final ajustado, sin vislumbrar un claro ganador. Sin embargo, el conteo rápido realizado por la ONG Súmate, a partir de una muestra estadística de mesas auditadas, arroja una diferencia entre el NO y el SI de más del 8%, una diferencia que implica más de 800.000 votos (*15*). Este resultado apunta en la dirección de una manipulación del recuento de votos. La posibilidad de que el CNE intentara maquillar los resultados, en una situación con poco margen de maniobra, no puede ser descartada. Si fuera el caso, al oficialismo no le quedó más remedio que aceptar la victoria del NO, so pena de desencadenar una crisis de incalculables consecuencias. La gran movilización de amplios sectores de la oposición y su presencia en los centros de votación hizo difícil que los resultados finales pudieran ser alterados. Además, no había consenso sobre la reforma en las filas del oficialismo, algo que incidió en la desmovilización de muchos chavistas y repercutió en la campaña que aquéllos desplegaron. Por último, según la información que ha trascendido, las Fuerzas Armadas hicieron respetar la victoria de la oposición.

En las elecciones parlamentarias de 2010 el gobierno recurrió a otra estrategia. Ante un previsible deterioro de los apoyos electorales, del que daban cuenta las encuestas, optó por una reforma electoral. Conforme al nuevo sistema electoral denominado "paralelo", el porcentaje de diputados a elegirse en circunscripciones (pluri) nominales aumentó, en promedio nacional, del 60% a cerca del 70%. Además, la reforma legalizó la práctica de las "morochas" ("gemelos", según el modismo venezolano), mediante la que el partido de gobierno había sido claramente sobrerrepresentado desde las elecciones regionales de 2004. Según esta práctica, un partido o coalición creaba otra organización ad hoc (morocha) que presentaba una lista en los circuitos electorales. Al tener esta nueva organización un estatus formal independiente, los partidos evitaban que se les restara los escaños obtenidos mediante el sistema pluralidad-mayoría. En definitiva, en el nuevo sistema electoral la elección de los postulados por lista se disocia de los candidatos uninominales. De ahí que los candidatos oficialistas pueden sacar partido de la gran popularidad de su líder. Sin lugar a dudas, la reforma es contraria al principio constitucional de la personalización del sufragio en distritos plurinominales y la representación proporcional (*16*). Finalmente, hubo modificaciones en las circunscripciones electorales de varios estados recurriéndose a un descarado *gerrymandering* como se puso de manifiesto en la doble estrategia utilizada: aislar/concentrar zonas que en el pasado habían votado en contra del Gobierno, y unión de áreas con un comportamiento electoral favorable al Ejecutivo con otras históricamente opositoras que en el nuevo diseño de los distritos quedaban "diluidas". Sin duda, el oficialismo salió muy beneficiado de la reforma electoral: aunque los partidos de oposición en su conjunto obtuvieron la mayoría de los votos, el partido gobernante logró la mayoría absoluta de escaños en la Asamblea Nacional.

**4. Estimación de resultados bajo hipótesis de fraude.** La metodología discutida no permite detectar mesas que hayan sido afectadas por irregularidades si estas no generan valores de $Z$ atípicos. Es posible que el porcentaje de mesas afectadas por las irregularidades sea mayor del que podemos detectar. Denotemos este porcentaje (desconocido) por $\beta \times 100\%$. El caso límite $\beta = 1$ indicaría que todas las mesas han sido afectadas, $\beta = 0.5$ que sólo el 50% de ellas, etc. Denotemos por $\varepsilon$ la diferencia (también desconocida) entre (a) la razón de votos a favor y votos válidos en las $\beta K$ mesas afectadas y (b) la misma razón si las mesas no hubiesen sido afectas. Sea $\rho$ la *verdadera* razón entre votos favorables y votos válidos en el total de mesas; sólo conocemos la razón poblacional que arroja el conteo automatizado de votos y que hemos denotado por $R$. En caso de fraude, si conociéramos los valores de $\beta$ y $\varepsilon$ podríamos estimar $\rho$ por $R - \beta\varepsilon$. Si bien no podemos estimar $\beta$, sí podemos estimar $\varepsilon$. Si $H3$ es cierta, las mesas con valores atípicos de $Z$ es una muestra de mesas afectadas por irregularidades y por ende la diferencia entre la razón en este conjunto de mesas y la razón poblacional es un estimador natural de $\varepsilon$. En términos de la notación que hemos introducido: sea $\kappa$ el número de mesas que tienen $Z$-score atípicos, por ejemplo menor que $-3.9$ o mayor que $+3.9$ (hay entre 101 y 326 outliers de esta clase, dependiendo del estudio de caso, ¡entre 34 y 60 veces más de lo esperado!). El estimador de $\varepsilon$ que comentamos es entonces $r_\kappa - R$. Consecuentemente, un estimador plug-in de $\rho$ es $\rho_\beta = R - \beta(r_\kappa - R)$. Calculamos este estimador para las elecciones y referéndums en lo que no hayamos descartado la hipótesis de fraude. Para cada caso consideraremos distintos escenarios posibles, haciendo variar $\beta$, y los comparamos con los resultados oficiales y otros datos disponibles reportados en las elecciones y referéndums en los que hemos aceptado la alternativa $H3$.

El referéndum revocatorio del 2004 ha sido ampliamente debatido (*6*). A diferencia del resto de los estudios de caso que hemos considerado, en él sólo se auditaron 150 mesas y está demostrado que la muestra no es ni representativa ni aleatoria (*17*). Adicionalmente, existen evidencias de que el conteo de votos pudo ser alterado en un alto porcentaje de mesas en el centro de totalización (*18*). Considerando escenarios extremos en que se alteraron entre el 70% y el 100% de las mesas automatizadas ($0.7 < \beta < 1$), nuestro estimador de $\rho$ oscila alrededor de 0.49 y 0.51, apuntando a un empate técnico, lo que sugiere que las irregularidades pudieron ampliar una victoria del oficialismo pero difícilmente determinaron al ganador. Algunos exit polls (*19*) sugieren que el resultado fue 61% a favor de la oposición y no 59% a favor del oficialismo, tal como informó el CNE, pero la confiabilidad de estas estimaciones es cuestionable (*6*). Lamentablemente, al menos en el caso venezolano, existen encuestas preelectorales y exit polls para todos los gustos.

A diferencia del referéndum del 2004, en las elecciones presidenciales del 2006 se auditaron aproximadamente el 54% de las mesas y se ampliaron los controles del sistema automatizado. Con estas garantías la oposición regresó a la contiende electoral, después de haber llamado a la abstención en las elecciones parlamentarios del 2005. Existe consenso en que la discrepancia arrojada por las auditorías no es significativa, descartando la hipótesis de fraude electrónico pero sigue siendo una incógnita la magnitud en la que otras irregularidades afectaron los resultados. Ni siquiera considerando el escenario límite, en el que todas las mesas resultaron afectadas por las irregularidades, podemos obtener estimaciones en el que se invierten los resultados. De hecho, $\rho_1 = 0.56$, aproximadamente un 8% por debajo de la proporción reportada por el CNE (62,84%). A lo sumo podemos apoyar la declaración del candidato perdedor (Manuel Rosales), quien reconoció la derrota con un margen más estrecho que el publicado.

El referéndum de la reforma constitucional del 2007 fue discutido en detalle en la sección anterior. A diferencia de las elecciones del 2006, las auditorías sí reflejaron diferencias importantes entre los votos emitidos y auditados. Conforme al análisis realizado por Súmate, la diferencia a favor de la oposición fue superior al 8%. Nuestras estimaciones son compatibles con este escenario para valores de $\beta$ cercanos a 0.5, sugiriendo la manipulación de resultados electorales para disimular una clara derrota del oficialismo, distinta al cuasi empate (50.7% versus 49.3%) decretado por el CNE.

La oposición reconoció con profundas reservas el triunfo del oficialismo en el referéndum de la enmienda constitucional del 2009, donde se aprobó la postulación del presidente Chávez a la reelección presidencial para el 2012. Al igual que en el 2006, las auditorías coincidieron con los resultados oficiales. Sin embargo, denunciaron una vez más las diversas irregularidades ya mencionadas, sin poder precisar el impacto de las mismas. Asumiendo valores de $\beta$ entre 0.25 y 0.45, obtenemos estimaciones de $\rho$ que oscilan entre 0.49 y 0.51, un empate técnico. Qué pasó en esas elecciones es una incógnita que nuestro análisis no puede develar. Muy posiblemente el resultado oficial del 55% a favor de la enmienda haya sobrestimado la opinión del electorado pero ciertamente no podemos concluir que las irregularidades determinaron los resultados.

En resumen, si bien no podemos descartar la hipótesis de fraude en comicios administrados por el actual árbitro electoral, difícilmente se ha perpetrado alguno que haya sido determinante en los resultados (*20*). De existir, posiblemente sólo haya ampliado la diferencia de los resultados a favor del oficialismo y acortado el margen con el que ganó la oposición en el 2007. ¿Qué pasará en las elecciones presidenciales del 2012 si los resultados son tan ajustados como algunas encuestas reconocidas vaticinan (*21*)? Nuestro estudio de casos sugiere que en este escenario la paleta de irregularidades podría voltear una elección a favor de Chávez, si es que este las llegase a perder.

**Referencias y Notas:**

1. The Carter Center (2004). Resumen ejecutivo del informe integral. Disponible online en http://www.cartercenter.org/documents/1839.pdf

2. De los cinco rectores del CNE sólo uno (Vicente Díaz) fue postulado por la oposición. El resto o han sido postulados por el partido de gobierno o han ejercido cargos públicos durante la presidencia de Chávez.

3. Ehremberg, R. (2012). Election night numbers can signal fraud. *ScienceNews* 181:4:16.

4. López-Pintor, R. Assessing electoral fraud in new democracies: a basic conceptual framework. IFES Conference Paper, February 2010.

5. CSI (Crime Scene Investigation) es una famosa serie de televisión de la cadena CBS.

6. Jiménez, R. (2011). Forensic analysis of the Venezuelan recall referendum. *Statist. Sci.* 26:4:564–583.

7. Levin, I., Cohn, G. A., Ordeshook, P. C. and Alvarez, R. M. Detecting voter fraud in an electronic voting context: an analysis of the unlimited reelection vote in Venezuela. EVT/WOTE'09 Proceedings of the 2009 conference on Electronic voting technology/workshop on trustworthy elections.

8. Myakgov, M., Ordeshook, P.C. and Shaikin D. (2009). *The Forensics of Election Fraud.* Cambridge University Press, New York.

9. Klimek, P., Yegorov, Y., Hanel, R., Thurner, S. (2012). It's not the voting that's democracy, it's the counting: Statistical detection of systematic election irregularities. arXiv:1201.3087v1.

10. Mebane, W. (2008). Election forensics: The Second-digit Benford's Law Test and recent American presidecial elections. In Election Fraud: Detecting and Deterring Electoral Manipulation. (R. M. Alvarez, T. E. Hall and S. D. Hyde, eds.) 162–181. Brooking Press, Washington, DC.

11. Pericchi, L. and Torres, D. (2011). Quick anomaly detection by the Newcomb–Benford Law, with applications to electoral processes data from the USA, Puerto Rico and Venezuela. *Statist. Sci.* 26:4:513–527.

12. Deckert, J., Myagkov, M. and Ordeshook, P. C. (2010). The irrelevance of Benford's Law for detecting fraud in elections. *Political Analysis,* 19:3:245-268.

13. McCoy, J. (1999). Chávez and the end of "Partyarchy" in Venezuela. *Journal of Democracy,* 10:3:64–77.

14. $S_k^2$ es una pequeña variación del estimador usado en (*6*), donde escalamos con la media poblacional en lugar de con la media muestral $\mu_k$. Teoría sobre estimadores de la razón puede encontrarse en Lohr, S. (2004). *Sampling: Design and Analysis, 2nd ed*. Brooks/Cole, Boston, MA.

15. Súmate (2008). *Informe Súmate. Referéndum sobre proyecto de reforma constitucional. Reporte de observación electoral.* Available online at http://www.sumate.org.

16. Hidalgo, M. (2011). The 2010 legislative elections in Venezuela. *Electoral Studies*, 30:4:872–875.

17. Hausmann, R. and Rigobón, R. (2011). In search of the black swan: Analysis of the statistical evidence of fraud in Venezuela. *Statist. Sci.* 26:4:543–563.

18. Martín, I. (2011). 2004 Venezuelan presidential recall referendum (2004 PRR): A statistical analysis from the point of view of data transmission by electronic voting machines. *Statist. Sci.* 26:4:528–542.

19. Prado, R. and Sansó, B. (2011). The 2004 Venezuelan presidential recall referendum: Discrepancies between two exit polls and official results. *Statist. Sci.* 26:4:502–512.

20. Hidalgo, M. (2009). Hugo Chávez's "Petro-socialism". *Journal of Democracy*, 20:2:78–92.

21. Sagarzazu, I. (2012).Venezuelan Pollsters, their Records and the 2012 Race-August Update. Disponible online en http://venezuelablog.tumblr.com/

**Figuras:**

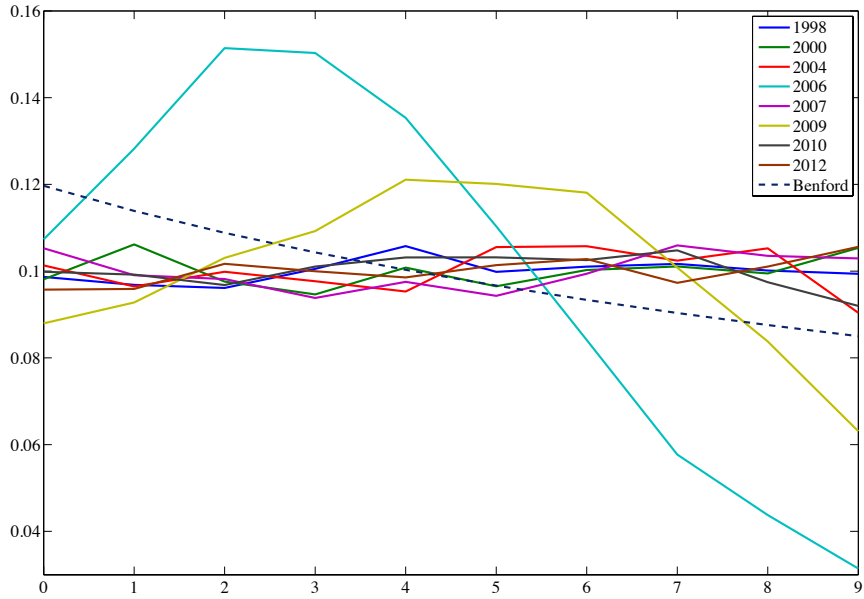

**Fig. 1.** Ley de Benford para el segundo dígito significativo vs. frecuencias observadas del número de abstenciones y votos nulos por mesa.

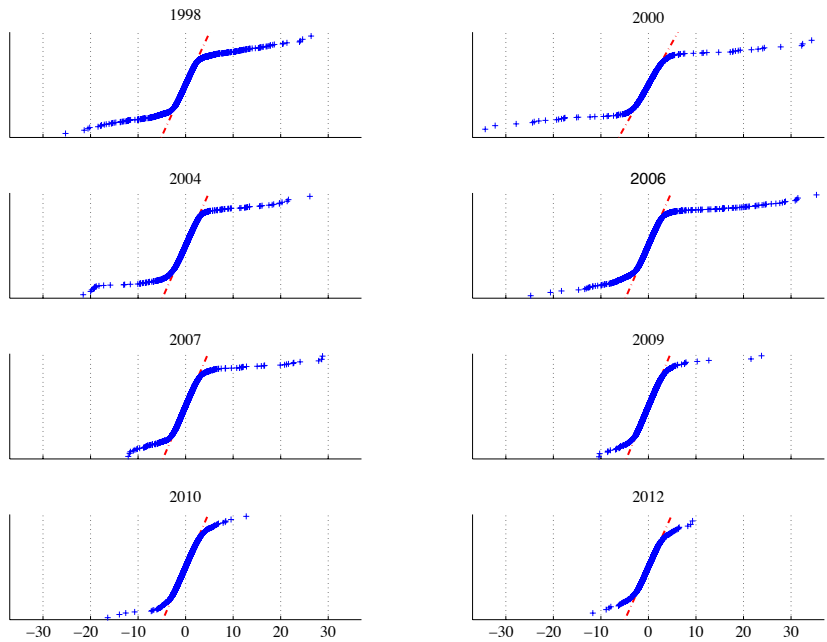

**Fig. 2.** Gráficos de probabilidad normal de $Z$-scores asociados al número de abstenciones y votos nulos (+ = valores observados y – – = valores esperados).

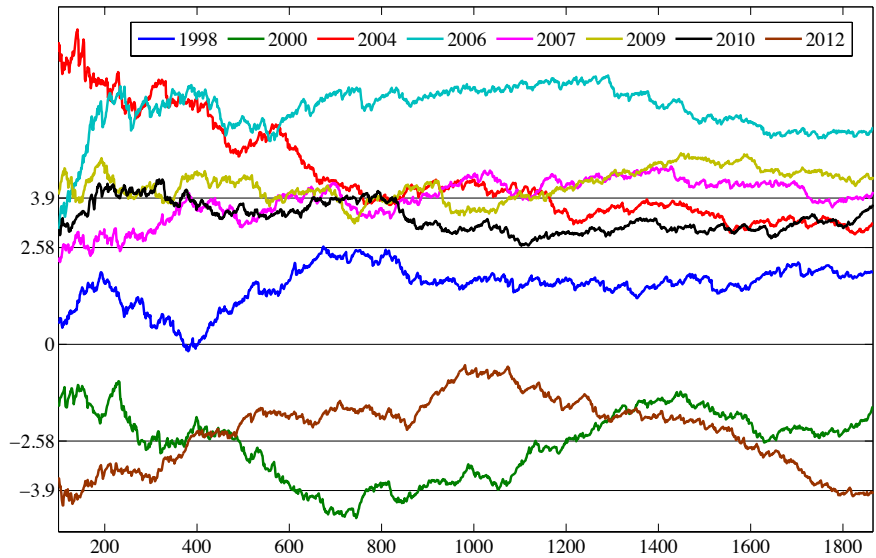

**Fig 3.** $\zeta_k$ versus $k$, para $100 < k < 1865$. El límite superior corresponde a la mitad del total de mesas bajo estudio del caso con menos mesas (elecciones del 2000).

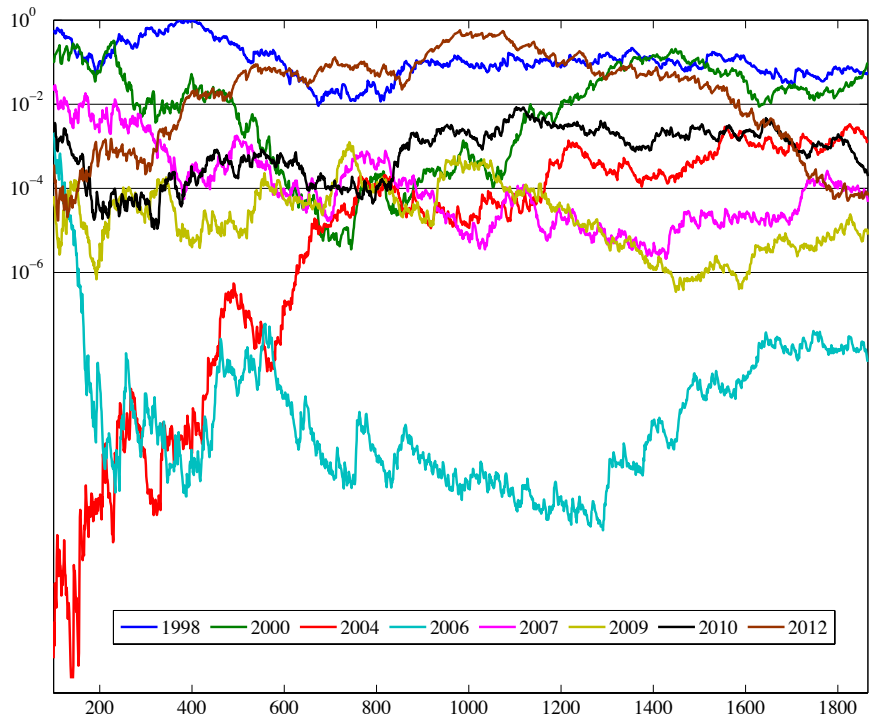

**Fig 4.** $p$-valores $P(|N(0,1)| > \zeta_k)$ en escala logarítmica para $100 < k < 1865$.

23